\documentstyle[12pt]{article}

\begin{document}

\begin{flushright}
LBL-35269
\end{flushright}

\begin{center}
{\bf
Comment on: ``Transverse-Mass \mbox{\boldmath $M_\perp$}
Dependence of Dilepton Emission from \\
\hspace{2.55cm} Preequilibrium and Quark-Gluon Plasma in High Energy\\
\hspace{-3.5cm} Nucleus-Nucleus Collisions"}

\end{center}

\vspace{0.5cm}
\begin{center}
{M. Asakawa\\
{\it Nuclear Science Division, Lawrence Berkeley Laboratory\\
 University of California, Berkeley, California 94720}}
\end{center}

\vspace{1cm}

In a recent Letter \cite{gmt}, Geiger presents calculations of the dilepton
emission from the early stage of ultrarelativistic heavy ion collisions using
the parton cascade model (PCM). He shows that the $M_\perp$
scaling \cite{mt85} found in Ref. \cite{ak93} is not observed. In this Comment,
we point out that this is largely due to a defect in the PCM.

The PCM is based on the perturbative QCD and inevitably contains two infrared
cutoffs. In particular, it has the lower limit of the timelike virtuality of
partons, $\mu_0=1$GeV. As a result, each parton has an effective mass larger
than 1GeV. During the nuclear collision, many partons acquire large timelike
virtuality and it decreases gradually to $\mu_0$. In the PCM, once the
virtuality reaches $\mu_0$, it is kept unchanged. This is expected to happen
typically by $\tau \simeq 1/\mu_0 = {\rm 0.2fm} \equiv \tau_{cut}$,
where the proper time $\tau$ is defined as
$\tau = {\rm sgn}(t-t_0 )\sqrt{(t-t_0 )^2 - z^2}$ \cite{gmt}
and in the center of mass frame, at $z=0$, the maximum density is achieved at
$\tau =0$. In reality, the virtuality of partons continues to decrease below
$\mu_0$ as time increases according to the uncertainty principle.
Therefore, the partons in the PCM have inappropriately large virtualities
after $\tau \simeq \tau_{cut}$.

This unphysical virtuality drastically affects the dilepton production,
especially that from $\bar{q}q$ annihilations with $M$ comparable to
$\sim \! 2\mu_0$ or less, where $M$ is the invariant mass of the dilepton.
A remarkable feature in Fig. 3 of Ref. \cite{gk}, on which the formalism of
Ref. \cite{gmt} is based, is strong suppression of the dilepton yield from
$\bar{q}q$ annihilations below $M\simeq2.8$GeV. At a glance, this might look
like a non-equilibrium effect. This is, however, not the case. Note that
schematic calculations of the dilepton emission from the preequilibrium
stage with on-shell quarks show enhancement of the dilepton production from
the $\bar{q}q$ process at small $M$ \cite{ma91}. Since this is a QED process
and there is no divergence in the process, the other cutoff in the PCM,
$p_{\perp cut}$, does not play a role here. This suppression is due to
phase space suppression by the large effective quark mass.
In this PCM calculation, there is no $\bar{q}q$ contribution below $M=2$GeV,
whereas in conventional calculations $\bar{q}q$ annihilations dominate this
region. From Fig. 3 of Ref. \cite{gk}, it is natural to expect that the
$\bar{q}{q}$ contribution would be  dominant also at
$M${\Large\lower3pt\hbox{$\,\buildrel <\over{\mbox{\normalsize $\sim$}}$}}
3GeV if it were not excessively suppressed by the large unphysical effective
quark mass.

According to Fig. 3 of Ref. \cite{gmt}, most of the $M_\perp$ scaling breaking
dileptons are created after $\tau=0.5$fm, when the PCM has too large effective
quark masses.  In the PCM local thermalization is achieved as early as
$\tau\simeq 0.3$fm \cite{gk2} and the transverse expansion does not establish
in the early stage of the collisions. Therefore, neither the preequilibrium
state nor transverse expansion is the reason for the $M_\perp$ scaling
breaking. It is the dileptons from the bremsstrahlung that are responsible for
the $M_\perp$ scaling breaking. The bremsstrahlung contribution is dominant at
small $M$ in the PCM \cite{gk} and does not generally realize the
$M_\perp$ scaling. For the $M_\perp$ scaling, the Boltzmann approximation is
essential \cite{mt85}, in which diagrams with outgoing (anti)quarks or gluons
are neglected. It cannot be used for the bremsstrahlung.

Since, as aforesaid, the other conditions for the $M_\perp$ scaling are
approximately satisfied as early as $\tau\simeq 0.3$fm, if the virtuality of
the partons is appropriately treated and the unphysical artificial suppression
of the $\bar{q}q$ process at small $M$ is removed, the large $M_\perp$ scaling
breaking obtained in Ref. \cite{gmt} is expected to disappear.

In summary, the large $M_\perp$ scaling breaking reported in Ref. \cite{gmt}
is due to the use of the inappropriately large virtualities
$\geq \mu_0$ after $\tau\simeq \tau_{cut}$ and the consequent suppression
of $\bar{q}q$ annihilations at low $M$, and is not physical. It is not
justified to use the present version of the PCM to calculate the dilepton
yield and confirm the $M_\perp$ scaling in the region of $M_\perp \sim 2-3$GeV.

We thank H. Heiselberg for discussions. This work was supported by the
Director, Office of Energy Research, Office of High Energy and Nuclear Physics,
Divisions of High Energy Physics and Nuclear Physics of the U.S. Department of
Energy under Contract No. DE-AC03-76SF00098.

\bigskip
\noindent{PACS numbers: 25.75.+r, 12.38.Mh, 24.85.+p}


\end{document}